\documentstyle[twocolumn,seceq,epsf]{jpsj}

\def\runtitle{Spectral Flow and Quantum Theory of Dissipation}
\def\runauthor{Masahiko {\sc Hayashi}}

\title
{
Spectral Flow and Quantum Theory of Dissipation \\
in the Vortex Core of BCS superconductors
}

\author
{ 
Masahiko {\sc Hayashi}\footnote{E-mail: 
hayashi@cmt.is.tohoku.ac.jp}
}

\inst
{
Graduate School of Information Sciences, 
Tohoku University, Sendai 980-8579
}

\recdate
{
\today
}

\abst
{
The dissipation process in 
two dimensional BCS superconductor 
due to the quasiparticles 
in the core of moving vortex  
is studied from a quantum mechanical point of view, 
especially paying attention to the spectral flow 
of the bound state energies in the vortex core. 
In order to clarify the nature of the 
spectral flow, 
we performed a numerical study 
of a finite system 
and, by extending the 
analysis to include the effect of impurities, 
we discuss the quantum mechanical 
origin of the dissipation in the vortex core. 
}
\kword
{
superconductivity, vortex, spectral flow, dissipation
}

\begin{document}
\sloppy
\maketitle

\section{Introduction}
Recently the vortices in 
type-II superconductors 
in the quantum limit, 
where the level spacing of 
the Caroli-de Gennes-Matricon's (CdGM's) 
bound states in the vortex core, 
$\hbar \omega_{0} = \Delta_{0}^{2}/E_{F}$ 
($\Delta_{0}$ and $E_{F}$ are 
bulk energy gap and 
Fermi energy, respectively) \cite{CdGM}, is 
larger than the energy scale of thermal 
fluctuation, $k_{B} T$ 
($k_{B}$ and $T$ are Boltzmann constant 
and temperature, respectively), 
have been attracting much attention 
stimulated by the study of 
high $T_{c}$ oxide and organic 
superconductors. 
Especially in the moving vortices, 
the effects of bound states 
are believed to be essential 
in understanding the vortex dynamics. 
In order to treat this problem, 
an interesting idea has been proposed, 
that is the spectral flow
\cite{Volovik,MM,Stone}. 
This is based on the finding that 
the bound state energies in 
the vortex core change 
continuously
when the vortex is moving. 
Employing this idea, 
the nondissipative part of the vortex dynamics 
has been studied intensively,
although there still remains controversy 
\cite{AT,Ao}. 

On the other hands, it is also an 
interesting problem to see 
how the dissipative part of vortex dynamics 
is modified from the Bardeen-Stephen's 
classical theory \cite{BS} 
in the quantum limit. 
Several attempts have been made to 
clarify this point. 
Until now, however, most of the studies 
were focused on the transition between 
CdGM's bound states 
due to the interaction with impurities 
\cite{KK,GP,vO,FS,LO}. 
In this letter, we argue that 
the transition between CdGM's 
bound states is driven 
by the spectral flow 
although it is significantly 
modified by the impurities. 
This point of view provides us 
with some new knowledge about 
the dynamics of quasiparticles 
in the core of moving vortices. 

In this letter, in order to clarify the nature 
of the spectral flow, we first perform 
a finite size study of 
the energy levels of quasiparticles 
in the presence of vortices. 
We consider a ring made from a strip of  
two dimensional (2D) BCS superconductor 
and concentrate on the case of single vortex 
in the system. 
The numerical evaluation of the eigenvalues 
of Bogoliubov-de Gennes equation 
clearly shows a spectral flow behavior. 
It is also shown that there are two kinds of 
spectral flow; 
one moving from the upper side to the lower 
side of the gap
and the other moving in the opposite direction, 
which we call 
$\varepsilon_{+}$- and
$\varepsilon_{-}$-branch, 
respectively. 

Based on these facts, we argue that 
the transition of an electron from 
$\varepsilon_{-}$- to $\varepsilon_{+}$-branch, 
which occurs due to 
impurities, gives the 
microscopic description of the quasiparticle scattering, 
{\it i.e.} the dissipation in the vortex core. 
This process is studied from a 
quantum mechanical point of view employing
Landau-Zener tunneling theory and 
it is clarified when the Bardeen-Stephen type 
description is valid. 

This letter is organized as follows: 
In Sec. \ref{BdGeq} we describe our 
model and basic scheme for 
the numerical calculation. 
Then the numerical result
is introduced, based on which the nature of the 
spectral flow is discussed. 
In Sec. \ref{imp} the effect of impurities 
is considered and the mechanism of the 
dissipation in the vortex core is discussed. 
In Sec. \ref{disc} we discuss the relation between 
our results and the preceding studies. 
Some further problems are also mentioned. 
In Sec. \ref{sum} we summarize our results.  

\section{Bogoliubov-de Gennes equation}
\label{BdGeq}

\begin{figure}
\begin{center}
{\parbox{4cm} 
{\epsfysize=4cm \epsfbox{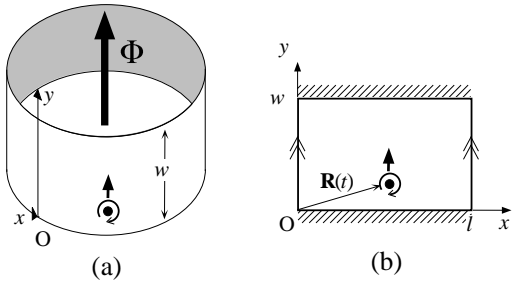}}}
\caption{
(a) Picture of the model system 
which we consider 
in this letter. Magnetic flux $\Phi$ is introduced to 
generate circulating supercurrent in the ring. 
A vortex (indicated by a dot) crosses the ring and 
causes decay of the supercurrent. 
(b) Equivalent area in $x$-$y$ plane. Lines 
at $x=0$ and $x=l$ are identified. The position of 
the vortex is denoted by ${\bf R}(t)$. }
\label{model}
\end{center}
\end{figure}

In this letter, we consider 
a ring made from a strip 
of 2D BCS superconductor (Fig. \ref{model}),
whose wave functions and 
energy spectrum of the quasiparticles 
are given by the Bogoliubov-de Gennes equation. 
The width and length 
of the strip are given by $w$ and $l$, 
respectively. 
Magnetic flux $\Phi$ is trapped in the cylinder to 
generate circulating supercurrent. 
We consider the case of single vortex
located at ${\bf R} (t) 
=(R_{x},R_{y})$ at time $t$. 

Bogoliubov-de Gennes equation 
is given by \cite{de_Gennes},
\begin{equation}
\left(
\begin{array}{cc}
\xi ({\bf r}) & \Delta({\bf r},t)\\
\Delta^*({\bf r},t) & - \xi^{*}({\bf r})
\end{array}
\right)
\cdot
\left(
\begin{array}{c}
u_{n}\\
v_{n}
\end{array}
\right)
= \epsilon_{n} \, 
\left(
\begin{array}{c}
u_{n}\\
v_{n}
\end{array}
\right), 
\label{BdG}
\end{equation}
where
\begin{equation}
\xi = -\frac{\hbar^2 }{2 m}
\left(\nabla - \frac{i e}{\hbar c}{\bf A}_{0}
\right)^{2} - E_{F}
+ V_{\rm imp}({\bf r})
\end{equation}
with $m$, $e$ and $c$ being the mass and charge 
($e < 0$) of electrons, and the velocity of light, 
respectively; 
${\bf A}_{0}=(\Phi/l)\, {\hat {\bf e}}_{x}$ with 
${\hat {\bf e}}_{x}$ being unit vector in $x$-direction; 
$u_{n}$ and $v_{n}$ are the 
wave functions of the quasiparticles 
and $\epsilon_{n}$ is their energy eigenvalue;
$V_{\rm imp}({\bf r})$ is the impurity potential. 
We disregard the fluctuation of the gauge field 
in this letter. 

\begin{figure}
\begin{center}
{\parbox{4cm} 
{\epsfysize=4cm \epsfbox{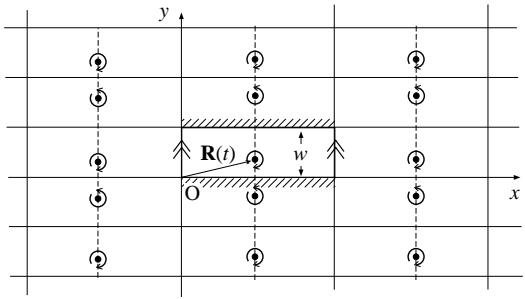}}}
\caption{Infinite number of mirror 
vortices generated by the boundary conditions 
of the system (in the middle).}
\label{mirror}
\end{center}
\end{figure}

First we solve Eq. (\ref{BdG}) with 
$V_{\rm imp}({\bf r})=0$
by introducing an approximate solution for 
$\Delta ({\bf r},t)$, 
instead of solving it selfconsistently. 
The approximate solution is constructed in the 
following way. 
It is well known that the condition 
of no outgoing supercurrent at the boundary
is satisfied by introducing 
mirror vortices. 
In the present case an infinite series 
of mirror vortices appears, 
as is seen from Fig. \ref{mirror}, 
which is considered to be two square vortex lattices 
with opposite flux overlapped with each other with 
a shift in $y$ direction. 
Therefore we can employ the order parameter 
of vortex lattice to construct 
an approximate order parameter 
for the present system. 
We use Eilenberger's order parameter 
\cite{Eilenberger} for 
the vortex lattice and apply a correction 
to the amplitude part in the sense of Clem's 
approximation \cite{Clem}, in order to make the amplitude 
constant except for the core region; 
\begin{eqnarray}
\phi_{\pm}({\bf r}) &=& 
{\rm e}^{-\pi ({\bar y} \mp {\bar R}_{y}+{\bar w})^{2}}
\vartheta_{3}\left({\bar x} + i ({\bar y} 
\mp {\bar R_{y}}+{\bar w}),
2 {\bar w} i\right)\nonumber\\
{\tilde \phi}_{\pm} &=& \phi_{\pm}/
\sqrt{\left|\phi_{\pm}\right|^{2}+c}\nonumber\\
\Delta &=& \Delta_{0}\,{\tilde \phi}_{+}\,
{\tilde \phi}_{-}^{*},
\end{eqnarray}
where $\vartheta_{3}(z,\tau)$ is the elliptic theta function; 
\lq bar\rq s denote 
lengths divided by $l$, {\it e.g.}, ${\bar x} = x/l$; 
$c$ is a positive constant which determines the radius of the 
core; $\Delta_{0}$ is the equilibrium energy gap 
outside the vortex core; 
${\tilde \phi}_{-}^{*}$ is the complex conjugate 
of ${\tilde \phi}_{-}$. 

We assumed parabolic dispersion for the Fermions 
and took into account only the states whose energy 
$E$ satisfies 
$E_{F} - \hbar \omega_{D} < E < E_{F} + \hbar \omega_{D}$ 
where $\hbar \omega_{D}$ is a cutoff parameter. 
The matrix elements of the Hamiltonian 
({\it i.e.}, the matrix of the 
left hand side of Eq. (\ref{BdG})) 
were calculated using this basis and 
energy eigenvalues were obtained after diagonalization. 

\begin{fullfigure}
\begin{center}
{\parbox{8cm}
{{\epsfysize=8cm {\epsfbox{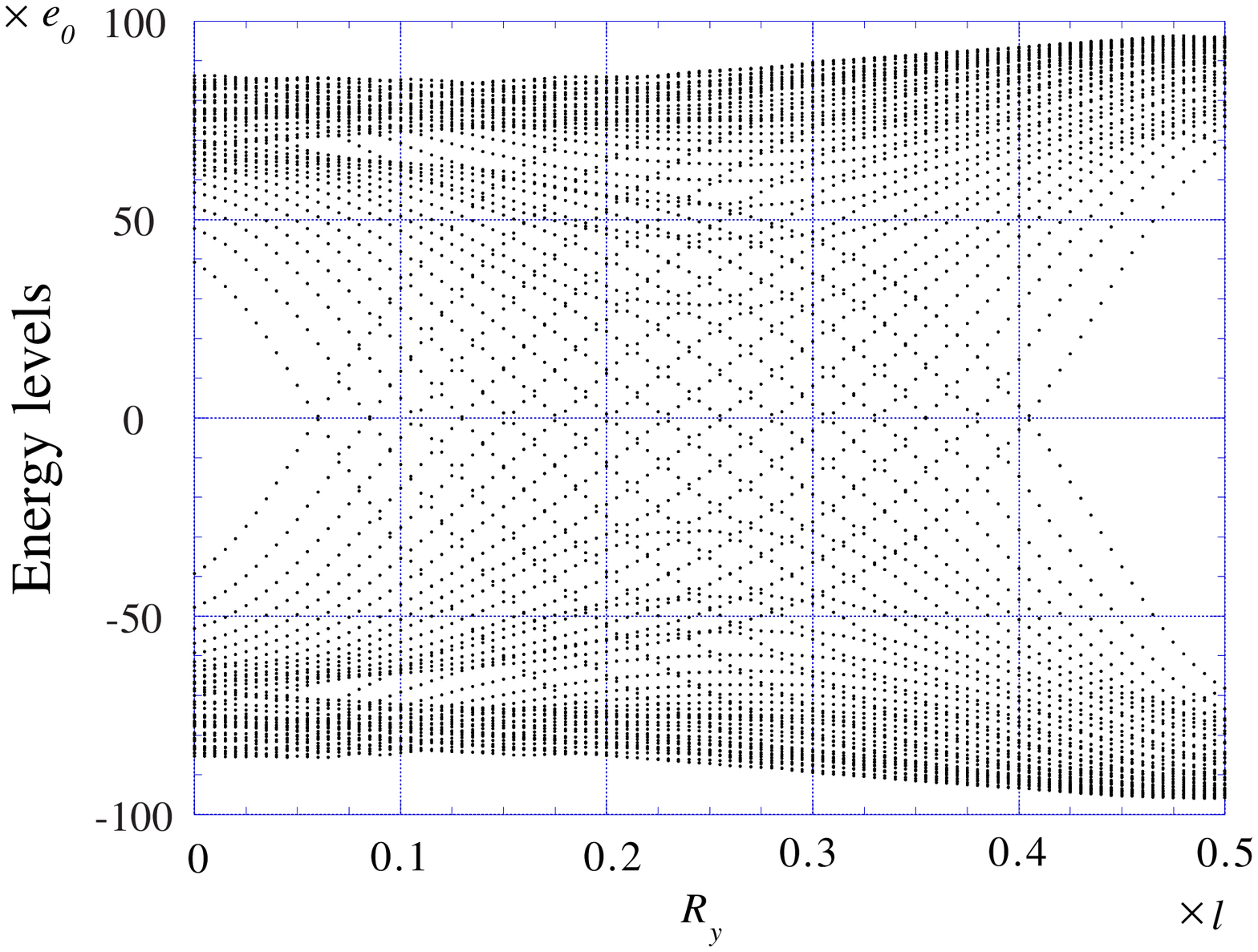}}}}}
\caption{Energy spectrum of quasiparticles
as a function of $R_{y}$. 
We can see the energy gap opening 
at $R_{y}=0$ and $R_{y} = w = 0.5 \times l$, 
and many energy level crossings in the middle. 
The crossing points are almost uniformly 
distributed in the range 
$0 < R_{y} < w$ except for the region which is 
approximately within the coherence length 
$\xi \sim 0.1 \times l$ from the boundaries. 
The gap at $R_{y} = w$ is smaller than 
$\Delta_{0}=10^{2} \times e_{0}$, because the average of the gap 
is suppressed by the effect of finite system size. 
The gap at $R_{y}=0$ is suppressed further, 
because the supercurrent 
is larger at $R_{y}=0$ than at $R_{y}=w$.  
Some of the energy levels 
outside the gap are not plotted. 
At some points, splitting of the crossing points  
is seen, which is due to the 
approximation and not essential.}
\label{spectrum}
\end{center}
\end{fullfigure}

In Fig. \ref{spectrum}, we show the result for 
the case of $w= 0.5 \times l$, 
$E_{F}=10^{3} \times e_{0}$, 
$\Delta_{0}=10^{2} \times e_{0}$ and 
$\hbar \omega_{D} = 300 \times e_{0}$ 
where $e_{0} = (\hbar^{2}/2 m)\, (2 \pi/l)^{2}$. 
Here $\xi = \hbar v_{F}/\Delta_{0} = 
\pi^{-1} l (E_{F}/\Delta_{0})\sqrt{e_{0}/E_{F}}$. 
Although the system is much smaller than the realistic one, 
our result shows a clear spectral flow behavior. 

Here we note a remarkable character of the spectral flow, 
namely the existence of many level crossings. 
This is the most essential process 
which leads to the dissipation in the vortex core, 
as we see in the next section. 
This character is understood in the following way: 
If $(u_{n},v_{n})$ is an eigenfunction 
with eigenvalue $\epsilon_{n}$, 
$(v^*_{n},u^*_{n})$ is also an eigenfunction 
with eigenvalue $- \epsilon_{n}$ \cite{de_Gennes}. 
Therefore the energy spectrum is 
symmetric with respect to 
the Fermi energy. 
This means that if there are 
several spectral flow branches which are 
flowing down crossing the Fermi energy, 
the same number of branches must be flowing up. 
These branches inevitably 
make crossing points. 

\section{Effect of impurity scattering}
\label{imp}

In this section we study 
the effect of impurities 
$V_{\rm imp}({\bf r})$. 
For this purpose 
it is useful here to 
look at the dynamics of quasiparticles 
in superconductor-normal metal-superconductor 
({\it S-N-S}) Josephson junction \cite{Beenakker,AB}, 
which shows a similar spectral flow behavior 
\cite{Stone}. 
In {\it S-N-S} Josephson junctions, 
the bound states are formed in $N$-region 
due to Andreev scattering at $S$-$N$ boundaries 
and, if we change 
the phase difference of two {\it S}'s, 
the energy levels of the bound states also 
change continuously \cite{Beenakker}. 
We can also see level crossings 
similar with our system. 
When a finite bias voltage is applied to the 
junction, the phase difference 
increases in time 
as is known from the Josephson's 
acceleration relation and 
the spectral flow occurs. 

A precise analysis of this system, when 
{\it N} region is a point contact, 
was given by Averin and Bardas \cite{AB}. 
They introduced a potential barrier 
in {\it N}-region and found that the 
scattering of quasiparticles 
by the barrier causes 
splitting of the crossing points of the 
spectral flow. 
The dominant scattering process 
is the backward scattering 
and the quasiparticles are scattered into the states 
with opposite momentum. 
They also found that the dynamics of the scattering is 
describable employing Landau-Zener tunneling theory. 

\begin{figure}
\begin{center}
{\parbox{8cm}
{\epsfysize=3.5cm \epsfbox{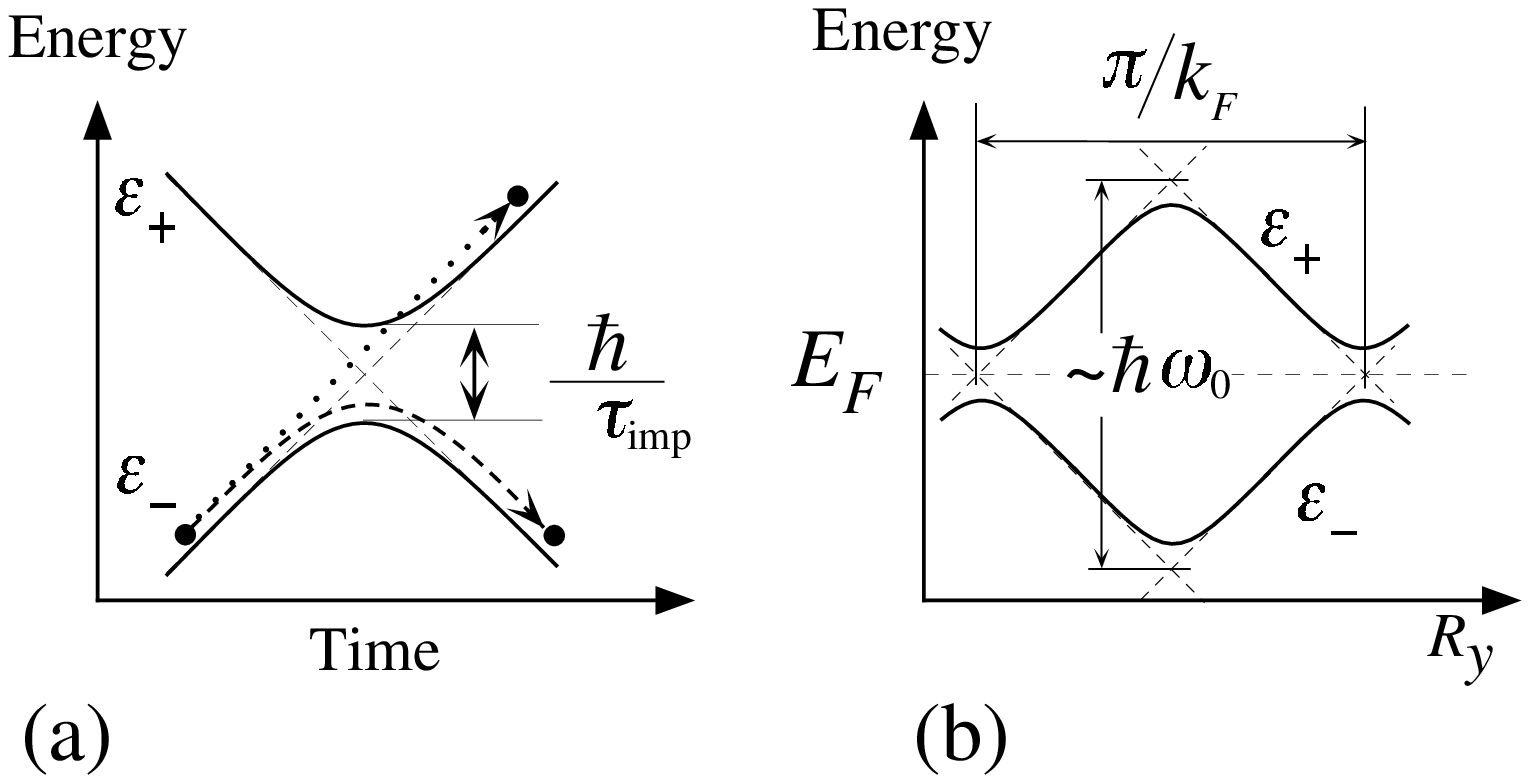}}}
\caption{(a) Level crossing and 
Landau-Zener tunneling process. 
(b) The energy and length scale near 
the crossing point.}
\label{Landau-Zener}
\end{center}
\end{figure}

From the analogy to the Josephson junctions, 
we can probably expect that the similar process 
occurs due to impurity scattering 
at each crossing points 
of Fig. \ref{spectrum}. 
We can approximately estimate the magnitude of 
the splitting as $\hbar / \tau_{\rm imp}$ where 
$\tau_{\rm imp}$ is 
the life time of quasiparticles in the normal state 
due to impurity scattering. 
When the vortex is moving with a velocity $v_{R}$, 
the scattering process can be 
described by the Landau-Zener tunneling theory. 
The time evolution of the system is shown 
in Fig. \ref{Landau-Zener} (a): 
If both of two crossing levels are vacant 
(or occupied) before crossing, the 
problem becomes rather trivial, 
since nothing special 
happens at the crossing point. 
However, if the lower level is occupied and 
the upper one is not, 
the final state changes depending on $v_{R}$. 
When $v_{R}$ is small, the particle 
follows the adiabatic change of the 
energy level, {\it i.e.}, the broken curve in 
Fig. \ref{Landau-Zener} (a). 
When $v_{R}$ is large, 
the particle has little time to interact with impurities 
and travels along the original levels 
without being scattered, 
{\it i.e.}, dotted line. 
The probability of the latter case 
is given by 
\begin{equation}
p=\exp\left\{- \frac{2 \pi (\hbar/\tau_{\rm imp})^{2}}
{\hbar\left(
{\dot{\varepsilon}}_{+}-{\dot{\varepsilon}}_{-}
\right)}\right\}
\label{LZ}
\end{equation}
where ${\dot{\varepsilon}}_{+}-{\dot{\varepsilon}}_{-}$
is the relative velocity of two levels in 
energy space, 
which is estimated as follows: 
At low temperatures $k_{B} T < \hbar \omega_{0}$, 
only the level crossings 
near the Fermi energy are relevant, 
which is depicted in 
Fig. \ref{Landau-Zener} (b). 
Firstly, the maximum of the spacing between two levels 
nearest to the Fermi energy is of the order of 
$\hbar \omega_{0}$, 
which approximately correspond 
to the CdGM's bound states with 
angular momentum $\mu = 1/2$ and $-1/2$. 
Secondly, in our system the wave number of electrons 
in $y$-direction is quantized by $\pi/w$. 
Therefore there exist $k_{F}/(\pi/w)$ channels, 
each of which is responsible for one pair of 
spectral flow 
($\varepsilon_{+}$ and $\varepsilon_{-}$). 
If we assume 
that the crossing points near the 
Fermi energy are distributed uniformly 
in $0 < y < w$ when $\xi \ll w$, we can estimate 
the separation between crossing points 
to be $\pi / k_{F}$, 
a half of the Fermi wave length. 
Although this is just an assumption at this stage, 
it seems to be a natural one, 
as one can see from Fig. \ref{spectrum}. 
Therefore, noting $R_{y} = v_{R} t$, we obtain
${\dot{\varepsilon}}_{+}-{\dot{\varepsilon}}_{-}
\simeq \hbar \omega_{0} k_{F} v_{R}/\pi$. 

We can see from Eq. (\ref{LZ}) 
that when $v_{R} k_{F} \tau_{\rm imp}^{2} 
\omega_{0} \ll 1$, the quasiparticle follows 
the change of the energy level adiabatically, 
which means that the quasiparticle is scattered 
by the impurity. 
In contrast with this, 
when $v_{R} k_{F} \tau_{\rm imp}^{2} 
\omega_{0} \gg 1$, the quasiparticle is not scattered 
but becomes a particle-hole pair after 
some level crossings, 
which costs approximately the energy of 
$\Delta_{0}$. 
This energy cost may probably 
suppress the vortex motion by 
acting as a nondissipative force. 
Further studies are needed to clarify the 
detail of this process, however. 

\begin{figure}
\begin{center}
{\parbox{8cm}
{\epsfysize=3.5cm \epsfbox{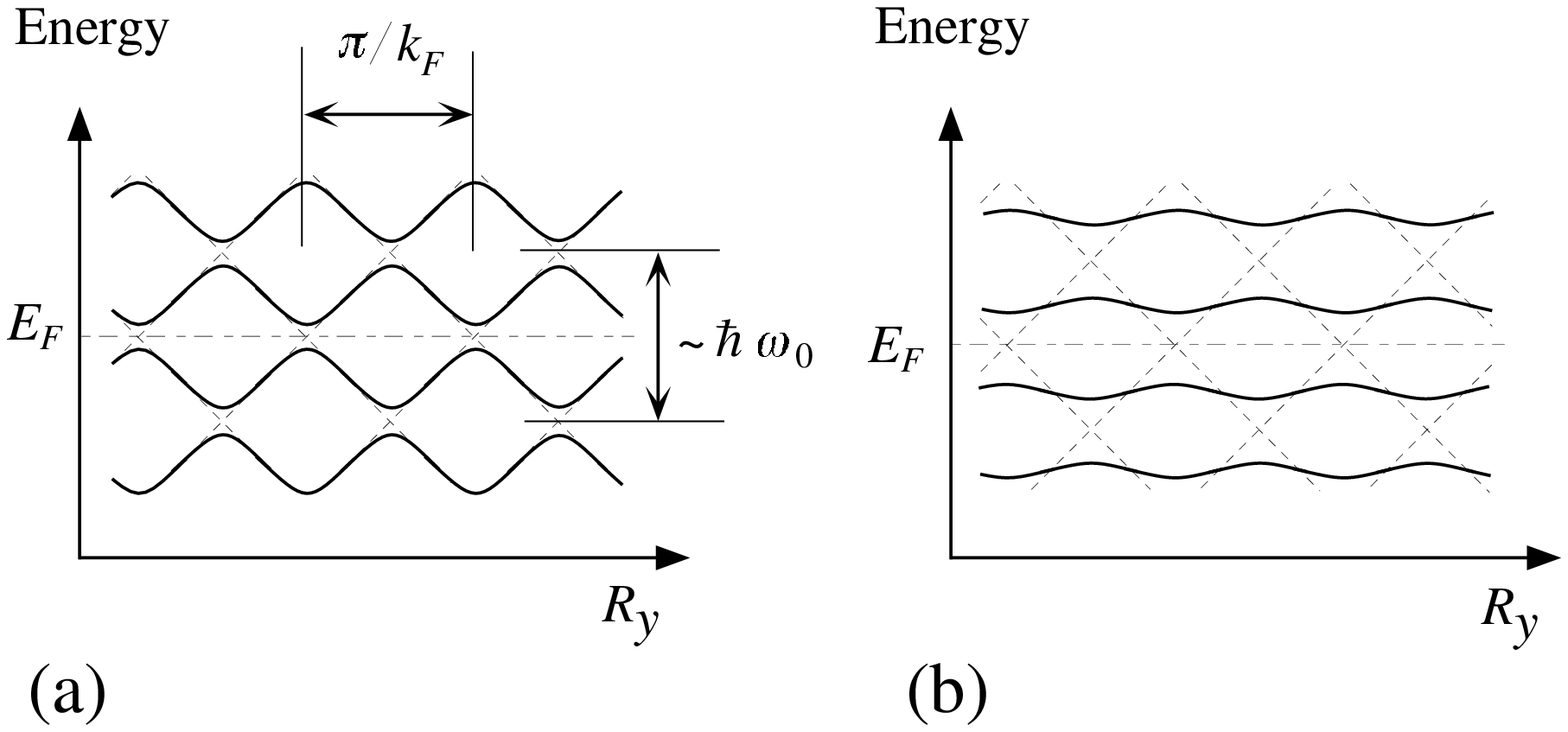}}}
\caption{Schematic view of the 
energy levels of the bound states 
in the case of 
(a) $\omega_{0} \gg 1 /\tau_{\rm imp}$ and 
(b) $\omega_{0} \ll 1 /\tau_{\rm imp}$.}
\label{energy_levels}
\end{center}
\end{figure}

It is also important to note that 
two different situations arise from 
the relation between the level spacing 
$\hbar \omega_{0}$ and the magnitude of splitting 
$\hbar /\tau_{\rm imp}$. 
When $\hbar \omega_{0} \gg \hbar /\tau_{\rm imp}$ 
the situation is depicted in 
Fig. \ref{energy_levels} (a). 
The energy levels oscillates with a period of 
a half of the Fermi wave length. 
In this case, 
a periodic potential for the vortex may appear due 
to the change of the quasiparticle energy. 
If we simply sum up the energies of 
the quasiparticles in the bound states, 
the amplitude of the periodic potential 
becomes the order of $\hbar \omega_{0}$ , 
which is not negligible 
at lower temperatures 
$k_{B} T < \hbar \omega_{0}$. 
On the contrary, 
when $\hbar \omega_{0} \ll \hbar /\tau_{\rm imp}$, 
the system recovers translational symmetry, 
as is seen from Fig. \ref{Landau-Zener} (b). 

The dissipation of Bardeen-Stephen type 
is, therefore, observed when 
$v_{R} k_{F} \tau_{\rm imp}^{2} 
\omega_{0} \ll 1$ and 
$\omega_{0} \ll 1 /\tau_{\rm imp}$ 
are satisfied at the same time. 

\section{Discussion}
\label{disc}

In the pioneering study, 
Caroli, de Gennes and Matricon
assumed that the vortex has rotational 
symmetry \cite{CdGM}. 
Therefore the angular momentum was taken to 
be a good quantum number. 
This, however, is not the case with our system, 
in which the boundaries break the 
rotational symmetry and cause mixing of 
different angular momentum states. 
This is the origin of the spectral 
flow in this letter. 
Volovik, Makhlin and Misirpashaev 
\cite{Volovik,MM}, 
by introducing the {\it laboratory frame},
obtained a similar spectral flow, 
which we consider is a realization of 
our finite size calculations in the bulk. 
The relation between two, however, is not clear 
at this stage. 

There are several works which have studied the 
dissipation in the vortex core from 
a quantum mechanical point of view,
by addressing the transition of 
quasiparticles between different CdGM's 
bound states due to impurity scattering 
\cite{vO,GP,FS,LO}. 
The largest difference of our study from 
those is that the transition of quasiparticles 
is driven by the spectral flow which exists even 
without impurity scattering. 
In our treatment, however, the effect of 
impurity scattering is only 
introduced by $\tau_{\rm imp}$ 
which is too simplified, especially in the case of 
small concentration of impurities. 
If we treat this more exactly, 
mesoscopic fluctuations may appear in 
the energy levels of Fig. \ref{energy_levels}, 
reflecting the character of 
individual impurities \cite{FS,LO}. 

In this letter we neglected the gauge fluctuation, 
namely the effect of finite penetration depth. 
Since the mirror vortices and their long-ranged 
phase modulation are important in our 
analysis, the effect of gauge fluctuation 
can be crucial. 
In order to clarify this point, 
we have to include the effect of gauge field 
into our Bogoliubov-de Gennes equation, 
which however needs further studies. 

In actual superconductors, three dimensionality is 
also important. 
In this case, the vortex become a line and 
each bound state in the vortex core obtains 
a continuous level 
corresponding to the motion along the line, 
which, of course, makes the problem much more
complicated. 
However, since the dissipation is 
mostly due to 
the backward scattering of quasiparticles, 
which is expected to occur between the states 
with the same momentum in 
the direction of the line, 
we may believe that the physical picture 
presented in this letter is still useful 
to understand the basic mechanism of the 
dissipation. 

\section{Summary}
\label{sum}
In this letter we studied the spectral flow 
and the origin of dissipation in 
the core of superconducting vortices 
from a quantum mechanical point of view. 
We found that several characteristic regimes are
defined by the material parameters 
and the vortex velocity. 
The dissipation of Bardeen-Stephen type occurs 
only when 
$v_{R} k_{F} \tau_{\rm imp}^{2} 
\omega_{0} \ll 1$ and 
$\omega_{0} \ll 1/\tau_{\rm imp}$ are satisfied. 
In other regimes the effect of the spectral flow 
does not result in simple dissipation but
can cause nondissipative 
forces as well. 

\section*{Acknowledgements}
The author would like to thank Prof. H. Fukuyama and 
Prof. H. Ebisawa for useful discussions.

\end{document}